\documentclass[11pt]{llncs}
\usepackage[margin=3.2cm]{geometry}
\pdfoutput=1

\usepackage{natbib}

\usepackage{epsfig}
\usepackage{epstopdf}
\usepackage{color}

\newcommand{\ket}[1]{{\left | {#1} \right\rangle}}

\newcommand\NP{${\cal NP}$}
\newcommand\Poly{${\cal P}$}
\newcommand\PSPACE{${\cal PSPACE}$}
\newcommand\BPP{${\cal BPP}$}
\newcommand\BQP{${\cal BQP}$}
\newcommand\QMA{${\cal QMA}$}
\newcommand\FPT{${\cal FPT}$}

\begin{document}

\title{\bf Computability and Complexity of Unconventional Computing Devices}
\author{
	Hajo Broersma
   	\inst{1}
\and
	Susan Stepney
   	\inst{2}
\and
	G\"oran Wendin
   	\inst{3}
}
\institute{
	Faculty of Electrical Engineering, Mathematics and Computer Science, \\ CTIT Institute for ICT Research, and \\  MESA+ Institute for Nanotechnology, \\  
	University of Twente, The Netherlands
\and
	Department of Computer Science, University of York, UK
\and
	Microtechnology and Nanoscience Department,
    Chalmers University of Technology, Gothenburg, SE-41296 Sweden
}

\maketitle

\begin{abstract}
We discuss some claims that certain UCOMP devices can perform hypercomputation (compute Turing-uncomputable functions) or perform super-Turing computation (solve \NP-complete problems in polynomial time).
We discover that all these claims rely on the provision of one or more unphysical resources.
\end{abstract}

\section{Introduction}

For many decades, Moore's Law \citep{Moore1969} gave us exponentially-increasing classical (digital) computing (CCOMP) power,
with a doubling time of around 18 months.
This cannot continue indefinitely, due to ultimate physical limits \citep{Lloyd2000}.
Well before then, more practical limits will slow this increase.
One such limit is power consumption.
With present efforts toward exa\-scale computing, the cost of raw electrical power may eventually be the limit to the computational power of digital machines: Information is physical, and electrical power scales linearly with computational power  (electrical power = number of bit flips per second times bit energy). Reducing the switching energy of a bit will alleviate the problem and push the limits to higher processing power, but the exponential scaling in time will win in the end. Programs that need exponential time will consequently need exponential electrical power.
Furthermore, there are problems that are worse than being hard for  CCOMP: they are (classically at least) \textit{undecidable} or \textit{uncomputable}, that is, impossible to solve.

CCOMP distinguishes three classes of problems of increasing difficulty \citep{Garey1979}:

\begin{enumerate}
\item Easy (tractable, feasible) problems: can be solved by a CCOMP machine, in polynomial time, $O(n^k)$, or better. 
\item Hard (intractable, infeasible) problems: take at least exponential time, $O(e^n)$, or exponential resources like memory, on a CCOMP machine. 
\item Impossible (undecidable, uncomputable) problems: cannot be solved by a CCOMP machine with any (finite) amount of time or memory resource.
\end{enumerate}

Unconventional Computing (UCOMP) \citep{UCOMP2009}
is a diverse field including a wealth of topics:
hypercomputation,
quantum computing (QCOMP),
optical computing,
analogue computing,
chemical computing,
reaction-diffusion systems,
molecular computing,
biocomputing,
embodied computing,
Avogadro-scale and
amorphous computing,
memcomputing,
self-organising computers,
and more.

One often hears that UCOMP paradigms  can provide solutions that go beyond the capabilities of CCOMP
 \citep{KonWen2014}. 
There is a long-held notion that some forms of UCOMP can provide tractable solutions to \NP-hard problems that take exponential resources (time and/or memory) for CCOMP machines to solve 
 \citep{Adleman1994,Lipton1995,Siegelmann1995,Ouyang1997,Copeland2004}, and the challenge to solve \NP-hard problems in polynomial time with finite resources is still actively explored \citep{Manea2007,Traversa2015a,Traversa2017}. 
Some go further, to propose UCOMP systems that can handle classically undecidable or uncomputable problems \citep{Cabessa2011,Copeland2011,Hogarth1992}.
 
Many of these analyses may be \textit{theoretically} sound, 
in that, if it were possible to implement the schemes, they would behave as claimed.
But, \textit{is} it possible to implement such schemes,
to build such a computer in the material world, under the constraints of the laws of physics?
Or are the hypothesised physical processes simply too hard, or impossible, to implement?

Key questions we discuss in this chapter are:

\begin{enumerate}

\item Can UCOMP provide solutions to classically undecidable problems? 

\item Can UCOMP provide more effective solutions to \NP-complete and \NP-hard problems?   

\item Are classical complexity classes and measures appropriate to any forms of UCOMP?

\item Which forms of UCOMP are clearly and easily amenable to characterisation and analysis by these? And why?

\item Are there forms of UCOMP where traditional complexity classes and measures are not appropriate, and what alternatives are then available?
\end{enumerate}

The idea that Nature is physical and does not effectively solve \NP-hard problems
does not seem to be generally recognised or accepted by the UCOMP community.
However, there is most likely no free lunch with UCOMP systems providing shortcuts, actually solving \NP-hard problems \citep{Aaronson2005}. 
The question is then, what is the real computational power of UCOMP machines: are there UCOMP solutions providing significant polynomial speed-up and energy savings, or more cost-effective solutions beyond the practical capability of CCOMP high performance computing, or different kinds of solutions for embodied problems, or something else?
This is the subject of the discussion in this chapter. 

In \S\ref{sec:problem} we discuss what it means to be a computational problem.
In \S\ref{sec:hyper} we discuss UCOMP and hypercomputation (computability) claims.
In \S\ref{sec:comp2} we recap the classical definitions of computational complexity.
In \S\ref{sec:qip} we discuss the power of various quantum computing approaches.
In \S\ref{sec:power} we discuss UCOMP and super-Turing computation  (complexity) claims, and the actual computational power of a variety of UCOMP paradigms.

\section{Computational problems and problem solving}
\label{sec:problem}
In the context of problem solving, the term complexity of a problem is
used to indicate the difficulty of solving that particular
problem, in many cases relative to the difficulty of solving other
problems. 
Two questions that need to be answered first are: what do we mean in this context by
\textit{problem} and by \textit{problem solving}?

\subsection{Difficulty}

Within the area of CCOMP,
solving a particular problem means developing an
algorithmic procedure that is able to produce a solution to that
problem. This assumes that the problem consists of a set of
instances, each of which can be encoded as an input to the algorithmic procedure, 
and that the algorithmic procedure then produces an output
that can be decoded into a solution for that instance of the problem.
This implies that being able to solve such types of problems means 
being able to write and install a computer program on a
digital device that, executed on an input representing any instance
of the problem produces an output that serves as a solution to that
particular instance. 

This leads to two natural questions:
\begin{itemize}
\item Does an algorithm exist for solving a particular problem?
This is a question of \textit{decidability} or \textit{computability}.
\item If such an algorithm does exist, how efficient is it at solving the problem?
This is a question of \textit{complexity}.
\end{itemize}

If such a computer program is available, it is natural to measure the
difficulty of solving the problem by the \textit{time} it takes the computer
program to come up with the solution. 
There are many issues with this measure.
For example, the execution time depends on the type and speed of the computer (processor), 
the type and size of the (encoding of the) instance, 
and on how smart the designed algorithmic procedure and its implementation were chosen.

In order to tackle some of these issues, 
the theory usually involves just the number of basic computational steps  in the algorithmic procedure,
and relates this to a function in the size of the instances. 
Upper bounds on the value of this function for the worst case instances are taken to indicate the relative
complexity of the problem when compared to other problems. 

Another natural question to ask is how much \textit{space} (memory) does the program need to use to solve the problem.
Again, the analyses abstract away from the complexities of actual computer memory (caches, RAM, discs, and so on), to an abstract concept of a unit of space.

This approach does not immediately say whether a more complex problem is intrinsically \textit{difficult},
nor whether the algorithmic procedure used is optimal or not in terms of the complexity. 
Identification of the least complex algorithm for problems is at the heart of the theory of computational complexity.

\subsection{Decision, optimisation, and counting problems}
\label{sec:decision}
There are different types of problems.
One distinction is based on the type of solutions. 

The main focus in the area of
computational complexity is on \textit{decision problems}, where the solution for each problem instance is
{\sc yes} or {\sc no}. 
The task is, given an arbitrary instance and a certain fixed property, to answer whether the given instance has that property.

Consider the travelling salesman problem (TSP).
An instance of TSP comprises a set of cities, together with the mutual distances between all pairs of cities.
A route is a permutation of the city list, corresponding to travelling through each city precisely once, returning to the starting city.
The length of the route is the sum of the distances between the cities as travelled. Given some value $x$ representing the length of the route,
TSP can be cast as a decision problem: is there a route that does not exceed $x$?
For a nice exposition on the many facets of TSP we refer the reader to \citet{Lawler1985}.

Consider the $k$-SAT (satisfiability) problem \citep{Garey1979}.
A formula (instance) in this problem involves any number $m$ of conjoined clauses, each comprising the disjunction of $k$ terms.
Each clause's $k$ terms are drawn from a total of $n$ Boolean literals, $b_1 \ldots b_n$, and their negations.  For example, a 3-SAT problem instance could be the formula $(b_1 \vee b_2 \vee b_3) \wedge (\neg b_2 \vee b_3 \vee b_5) \wedge (b_1 \vee \neg b_3 \vee b_4) \wedge (\neg b_1 \vee b_3 \vee \neg b_5)$,
which has $n=5$ and $m=4$.
$k$-SAT is a decision problem: is there an assignment of truth values to the $b_i$ that satisfies (makes true) the formula?

Decision problems differ from problems for which the solution is other than just {\sc yes} or {\sc no}. 
A large class of problems for which this is the case, is the class of so-called
\textit{optimisation problems}. For these problems, it is not sufficient to come up with solutions, but the
solutions are required to satisfy certain additional optimisation criteria. 
TSP can be cast as an optimisation problem: what is (the length of) a shortest route?

Another large class
of problems that are not decision problems, is the class of \textit{counting problems}. For these problems, the
solutions are numbers rather than {\sc yes} or {\sc no}. For TSP, one could, e.g., ask for the number of routes that are shorter than $x$, or for the number of different shortest routes.

Most optimisation and counting problems have decision counterparts (as is clear in the case of TSP above).
Such optimisation and counting problems are obviously at least as difficult to solve as their decision counterparts.

\subsection{Terminology}

We use the term \textit{hypercomputation} to refer to UCOMP models that can compute classically uncomputable functions (such as the Halting function, a total function that decides whether an arbitrary computer program halts on an arbitrary input).  This is sometimes referred to as computation that ``breaks the Turing barrier'' or is ``above the Turing limit'' (that is, the barrier to, or limit on, computability).

We use the term \textit{super-Turing computation} to refer to UCOMP models that can compute more efficiently (using exponentially fewer resources) than a Deterministic Turing Machine (DTM). 

The UCOMP literature is not consistent in its use of these terms.
Careful reading may be needed to determine if a particular claim is about computability or about complexity.

\section{A brief review of CCOMP Computability}
\label{sec:comp1}

\subsection{Undecidable problems, uncomputable functions}
Not all problems can be solved by an algorithmic procedure using a classical computer. 
It has been known since \citet{Turing1937} that there are \textit{undecidable} problems:
those for which there is provably no algorithmic procedure to produce the correct {\sc yes/no} answer.

The earliest example of such a problem is the \textit{Halting Problem}. In this problem, one has to write a computer program $H$ that takes as its input any computer program $P$ and input $I$, and outputs {\sc yes} if $P$ would eventually halt (terminate) when run on $I$, and outputs {\sc no} otherwise.
There is provably no such $H$.
Since then, many other
examples of such undecidable problems have been established.

For problems not cast as decision problems, but in terms of computing the value of a function defined in a finite number of well-defined steps,
there are \textit{uncomputable} functions.
Well-known examples include Kolmogorov complexity \citep{Li1997},
the Busy Beaver function \citep{Rado1962},
and Chaitin's omega halting probability \citep{Chaitin1975,Chaitin2012}.
Note that by function in the above we mean
a function on the natural numbers; such a function $F$ is \textit{(Turing) computable} if there is a Turing Machine that, on input $n$, halts and returns output $F(n)$. The use of Turing Machines here is not essential; there are many other models of computation that have the same computing power as Turing Machines.

The existence of (many) uncomputable functions of the above type follows from the fact that there are only \textit{countably} many Turing Machines, and thus only countably many computable functions, but there are uncountably many functions on the natural numbers.
Similarly, a set of natural numbers is said to be a \textit{computable set} if there is a Turing Machine that, given a number $n$, halts with output 1 if $n$ is in the set and halts with output 0 if $n$ is not in the set.
So for any set with an uncountable number of elements, most of its elements will be uncomputable.
Hence most subsets of the natural numbers are uncomputable.

Decision problems can be encoded as subset problems: encode the problem instance as a unique natural number; the {\sc yes} answers form a subset of these numbers; the decision problem becomes: is the number corresponding to the problem instance an element of the {\sc yes} set? Hence most decision problems are uncomputable, that is, undecidable.

These undecidable problems and uncomputable functions are hard to solve or compute in a very strong sense: within the
context of CCOMP it is simply impossible to solve or compute them.

\subsection{Oracles and advice}
\label{sec:oracle}

Computability is an all or nothing property (although whether a problem class is computable may itself be an uncomputable problem).
\textit{Oracles} can be used to add nuance to this property:
how much (uncomputable) help would be needed to make a problem computable?
Less powerful oracles can also be considered when investigating complexity:
how much oracular help is required to reduce the complexity of a problem class?

An oracle is an \textit{abstract} black box that can take an input question from a DTM and output the answer.  Oracles can be posited that provide answers to certain classes of problems, such as halting-problem oracles and \NP-problem oracles.  An oracle is usually deemed to provide its answer in one step. 
(See \S\ref{sec:PandNP} for a definition of classes \Poly\ and \NP.)

Oracles can be posited, and their consequent abilities investigated theoretically, but they cannot be implemented on a classical computer, since they provide computational power above that of a DTM.

More recently introduced complexity classes try to capture additional computational power provided by allowing \textit{advice} strings. 
An advice string is an extra input to a DTM that is allowed to depend on the length of the original input to the DTM, but not on the value of that input. A decision problem is in the complexity class \Poly$/f(n)$ if there is a DTM that solves the decision problem in polynomial time for any instance $x$ of size $n$ given an advice string of length $f(n)$ (not depending on $x$).

Trivially, any decision problem is in complexity class \Poly/exp.  If the input is of size $n$, then there are $O(2^n)$ possible input values $x$ of size $n$.
An exponentially large advice string can enumerate the $O(2^n)$ {\sc yes/no} answers to the decision problem as an exponentially large lookup table.

Advice strings can be posited, and their consequent abilities investigated theoretically.  Given an advice string, it can be implemented along with the DTM using its advice, since the string could be provided as an input to the DTM.  However, classically, the advice on that string would itself have to be computed somehow; if the string contains uncomputable advice, then classically it cannot exist to be provided to the DTM.

\subsection{Church-Turing thesis}
\label{sec:CTT}

The {\em Church-Turing Thesis} (CTT) states that ``every number or function that `would naturally be regarded as computable' can be calculated by a Turing Machine'' \citep{Copeland2015}. 
This is a statement about computability, in terms of a (digital) classical computer.

\citet{Vergis1986} reformulate this thesis in terms of analogue computers as:
``any analogue computer with finite resources can be simulated by a digital computer''.
This is a statement about computability: (finite) analogue computers do not increase what is computable over classical digital computers.

Hypercomputation seeks to discover approaches that can expand the range of computable functions beyond those computable by Turing Machines; it seeks to invalidate the CTT.

\section{Hypercomputation}
\label{sec:hyper}

\subsection{Undecidable problems determined physically?}

Hypercomputation is a diverse field with many ideas on how to compute classically uncomputable functions using physical and non-physical approaches.
One of the major proponents of hypercomputation is Jack Copeland  \citep{Copeland2004,Copeland2011,Copeland2015}.
\citet{Arkoudas2008} states:  
\begin{quote}
Copeland and others have argued that the CTT has been widely misunderstood by philosophers and cognitive scientists. In particular, they have claimed
that the CTT is in principle compatible with the existence of machines that compute functions
above the ``Turing limit'', and that empirical investigation is needed to determine the
``exact membership'' of the set of functions that are physically computable.
\end{quote}

\citet{Arkoudas2008} disputes this argument, and claims that it is a category error to suggest that what is computable can be studied empirically as a branch of physics, because computation involves an \textit{interpretation} or \textit{representation} component, which is not a concept of the physical sciences.  (See also \citet{Horsman2014}.)

\subsection{Accelerated Turing Machines}
An example of a theoretical hypercomputer is the Zeno Machine 
\citep{Potgieter2006}. A Zeno Machine is an Accelerated Turing Machine that takes $1/2^n$ units of time to perform its $n$-th step; thus, the first step takes $1/2$ units of time, the second takes $1/4$, the third $1/8$, and so on, so that after one unit of time, a countably infinite number of steps will have been performed. In this way, this machine formally solves the Halting Problem: is it halted at $t=1$?. 

Such a machine needs an exponentially growing bandwidth (energy spectrum) for operation, which is not a physically achievable resource. 

Any physical component of such a machine would either run up against relativistic limits, and be moving too fast, or quantum limits, as it becomes very small, or both.  The model implicitly relies on  Newtonian physics.

\subsection{General Relativistic Machines}

There are various models that use General Relativistic effects to allow the computer to experience a different (and infinite) proper time from the (finite) time that the observer experiences.
The best known of these is the Malament--Hogarth spacetime model \citep{Etesi2002,Hogarth1992}.
The underlying concept is that the computer is thrown into one of these spacetimes, where it can be observed externally.  If a computation does not halt, this can be determined in a finite time by the observer in the external reference frame, and so the set-up solves the Halting Problem.

This is an interesting branch of work, as it demonstrates clearly how the underlying laws of physics in the computer's material world can affect the reasoning used about possible computations.

However, there are several \textit{practical} issues with this set-up.
The computer has to be capable of running for an infinite time in its own reference frame.  Also, the ``tape'' (memory) of the computer needs to have the potential to be actually infinite, not merely unbounded.  It is not clear that such infinite time and infinite space can be physically realised.

\subsection{Real number computation}

A model of computation beyond the Turing limit has been formulated by \citet{Siegelmann1995}, involving neural networks with real-valued weights. 
\citet{Douglas2003,Douglas2013} provides a critical analysis. The problem is the usual one for analogue systems: ultimate lack of precision; {\it in the end one needs exponential resources}.
Analogue precision can be converted (by an ADC, Analogue-Digital Converter) into a corresponding digital range, which is effectively a memory requirement. \NP-problems (see later) require exponentially growing analogue precision, corresponding to a need for exponentially growing memory.
Hypercomputational problems (computability) correspond to a need for infinite precision.

Real number hypercomputation \citep{Blum2001} relies on physical systems being measurable to infinite precision.
The underlying argument appears to be: 
physicists model the physical world using real numbers; 
real numbers have infinite precision and so contain infinite information; 
hence physical systems have infinite information;
this information can be exploited to give hypercomputation.
There are two problems with this argument.

The first problem is that the argument confuses the model and the modelled physical reality.  Just because a quantity is modelled using a real number does not mean that the physical quantity faithfully implements those real numbers.
The real-number model is in some sense `richer' than the modelled reality;  it is this extra richness that is being exploited in the theoretical models of hypercomputation.
For example, consider Lotka--Volterra-style population models \citep{Wangersky1978}, where a real-valued variable is used to model the population number, which is in reality a discrete quantity: such models break down when the continuum approximation is no longer valid.  
Fluid dynamics has a continuous model, but in the physical world the fluid is made of particles, not a continuum, and so the model breaks down.
The Banach--Tarski paradox \citep{Wagon1985,Wapner2005} proves that it is possible to take a sphere, partition it into a finite number of pieces, and reassemble those pieces into two spheres each the same size as the original; the proof relies on properties of the reals that cannot be exploited to double a physical ball of material. 

Secondly, even if some physical quantity were to contain arbitrary precision, there is strong evidence that it takes an exponentially increasing time to extract each further digit of information (see section~\ref{sec:phys-oracle}).

\subsection{Using oracles; taking advice}

\citet{Cabessa2011} state: 

\begin{quote}
The computational power of recurrent neural networks is intimately related to the nature of their
synaptic weights. In particular, neural networks with static rational weights are known to be Turing
equivalent, and recurrent networks with static real weights were proved to be [hypercomputational]. Here, we
study the computational power of a more biologically-oriented model where the synaptic weights can
evolve rather than stay static. We prove that such evolving networks gain a [hypercomputational] power, equivalent to that of static real-weighted networks, regardless of whether their
synaptic weights are rational or real. These results suggest that evolution might play a crucial role in
the computational capabilities of neural networks.
\end{quote}

A proposed rational-number hypercomputer avoids the issue of infinite precision.
The set-up described by \citet{Cabessa2011}
is a neural network with rational, but changing (`evolving'), weights.
However, the changing weights are not computed by the network itself, nor are they provided by any kind of evolutionary feedback with a complex environment.
They are provided directly as input in the form of a sequence of increasing-precision rational numbers: that is, an advice string.

Any claim of hypercomputation achieved through the use of an oracle or advice needs to address the feasibility of implementing said oracle or advice.
These can provide hypercomputational power only if they are themselves Turing-uncomputable,

\subsection{Conclusion}

Hypercomputation models tend to rely on one or more of:
\begin{itemize}
\item known incorrect models of physics (usually Newtonian, ignoring relativistic and/or quantum effects)
\item physically-instantiated infinities (in time and/or space and/or some other physical resource)
\item physically accessible infinite precision from real numbers
\item Turing-uncomputable oracles or advice
\end{itemize}

There is currently no evidence that any of these essentially \textit{mathematical} hypercomputation models are physically realisable.  
Their study is interesting, however, because they illuminate the various relationships between computability and physical (as opposed to mathematical) constraints.  For example, they bring into focus an unconventional computational resource: precision.

Other facets of hypercomputation, of moving beyond computational paradigms other than Turing, are discussed in \citet{Stepney2009-hyper}.

\section{A brief review of CCOMP Complexity}
\label{sec:comp2}

We now move on to discussing super-Turing UCOMP models, which deal with computational complexity.
First we briefly review some concepts of classical complexity theory.
Further information can be found in any good textbook on computational complexity, such as \citet{Garey1979,Sipser1997}.

\subsection{Measuring complexity}
\label{sec:measure}
Complexity in the classical setting of digital computing is typically a mathematically calculated or proven property, rather than an empirically measured property, for two main reasons.

Firstly, complexity refers to asymptotic properties, as the problem sizes grow.  It would have to be measurable over arbitrarily large problem sizes to determine its asymptotic behaviour.  This is particularly challenging for many UCOMP devices (including quantum computers) that to date only exist as small prototypes that can handle only small problem instances.

Secondly, complexity is a worst case property: the complexity of a problem (or class) is the complexity of the hardest problem instance (or hardest problem in that class). 
Some instances can be easy, other instances hard.
If there are very few hard instances, these ``pathological'' instances may not be encountered during empirical sampling.

\subsection{Easy, polynomial time problems}

\citet{Edmonds1965} was the first to distinguish good and bad
algorithmic procedures for solving decidable problems. He coined the term \textit{good algorithm} for an
algorithmic procedure that produces the correct solution using a number of basic computational
steps that is bounded from above by a polynomial function in the instance size. 
He also
conjectured that there are decidable problems for which such good algorithms cannot be designed.
This conjecture is still open, although there is a lot of evidence for its validity.
We come back to this later.

The algorithms that Edmonds called good, are  nowadays usually referred to as \textit{polynomial} (time) algorithms,
or algorithms with a polynomial (time) complexity.
The corresponding problems are usually called polynomial(ly solvable) problems, but also referred to as easy, tractable, or feasible problems.

Define the function $f(n)$ to be the bound on an algorithm's number of basic computational steps in the worst case, for an instance of size $n$. 
Then for a polynomial (time) algorithm, $f(n)$ is
$O(n^k)$, meaning that there exists a positive integer $k$ and positive constants $c$ and $n_0$
such that
$f(n)\le c\cdot n^k$ for all $n\ge n_0$. 
The $n_0$ is included in the definition because
small problem instances do not determine the complexity: the complexity is  characterised by what happens for \textit{large} problem instances.

\subsection{Hard, exponential time problems}
Decidable decision problems for which no tractable (polynomial time) algorithm exists are called hard, intractable, or infeasible problems. 
These usually allow straightforward exponential algorithms: algorithmic procedures for solving them have worst case instances that take an exponential number of computational steps.
The function $f(n)$ that bounds the number of
computational steps, in the worst case for
an instance of size $n$, is $O(c^n)$, for some positive constant $c$.

Similar to the conjecture of \citet{Edmonds1965}, it is nowadays widely believed that there are decision problems for which the only possible algorithmic procedures for solving them have an exponential complexity.

\subsection{Complexity classes {\Poly} and \NP}
\label{sec:PandNP}
Based on the above distinction between easy and hard problems, the formally defined complexity class {\Poly} consists of all decision problems that
are tractable, that admit polynomial algorithms for solving them by a {\em classical Deterministic Turing Machine (DTM)}. 
A decision problem in {\Poly} is
also referred  to as a problem with a polynomial complexity, or simply as a polynomial problem.
The decision version of TSP described above has no known polynomial time algorithm to solve it.

Many textbooks, such as  \citet{Garey1979,Sipser1997},
provide a fundamental treatment of complexity class {\Poly} in terms of Turing Machines.
There, it is argued that the problems in {\Poly} are precisely those problems that can be encoded and
decided on a DTM within a number of transitions between states that is bounded by a polynomial
function in the size of the encoding (number of symbols) of the input.

The class \NP\ can be defined in terms of Turing Machines as  consisting of those problems that can be decided on a \textit{Non-deterministic Turing Machine (NTM)} in polynomial time.
A DTM has only one possible move at each step, determined by its state transition function along with its internal and tape state.
In contrast, an NTM has potentially several alternative moves available at each step,
and chooses one of these non-deterministically.
The computation succeeds if at least one sequence of possible choices succeeds.

There are alternative ways to consider the working of an NTM:
(i) it uses \textit{angelic non-determinism} and always makes the correct choice;
(ii) at each choice point, it `branches' into parallel machines, taking all possible paths (hence using exponential space).
An NTM can be implemented by serialising this branching approach \citep{Floyd1967}:
if it has chosen a path and discovers it is the wrong path, it `backtracks' and makes a different choice 
(hence potentially using exponential time).
Hence a DTM can compute anything an NTM can compute, although potentially exponentially slower.

An NTM can also be considered as an \textit{oracle machine}, a black box that provides candidate solutions for a specific class of problems (see section~\ref{sec:oracle}).
In the terminology used in  
textbooks like \citet{Garey1979,Sipser1997}, 
a decision problem is in the class \NP\ if, 
for any {\sc yes}-instance of the problem there is a candidate solution that can be checked by an algorithmic procedure in
polynomial time. So, instead of finding the correct answer for any instance in
polynomial time, it is only required to be able to verify the correctness of a candidate solution for the {\sc yes}-answer for any
{\sc yes}-instance in polynomial time. 

The decision version of TSP described above is in \NP: the relevant candidate solution is a suitable short route, and the length of that given route can be calculated in polynomial time and checked to be at most $x$.
Interpretation (i) of the NTM above has it `angelically' making the correct choice at each city; interpretation (ii) has it exploring all exponential number of possible paths in parallel.

It is clear that {\Poly} $\subseteq$ \NP: if a problem can be solved in polynomial time, it can certainly be checked in polynomial time.

It is widely believed that {\Poly} $\neq$ \NP\ (and that is the position we take in this chapter). Its proof (or disproof) is a fundamental open
problem within mathematics and theoretical computer science \citep{Aaronson2017}. Many decision problems have been shown to be in {\Poly}, but for even
more, such as TSP, it is not known whether they are in {\Poly} or not. 

\subsection{\NP-complete problems}
There are many decision problems for which the complexity status is currently unknown. 
To say at least something about their relative complexity, 
\citet{Cook1971} and 
\citet{Levin1973} developed useful machinery, which has led to the definition of the class of \NP-complete problems.

A problem in \NP\ is called \NP-\textit{complete} if it is the hardest of all problems in \NP, in the following sense.
Consider two problems $P$ and $Q$ that are both in \NP. Suppose that there exists a
polynomial reduction from $P$ to $Q$, that is, a
polynomial algorithm to transform
any instance $I$ of $P$ into an instance $J$ (of size bounded by a polynomial function in the size of $I$)
of $Q$ in such a way that $I$ is a {\sc yes}-instance of $P$ if and only if $J$ is a
{\sc yes}-instance of $Q$. 
Then any polynomial algorithm for solving $Q$ can be transformed into a polynomial algorithm for solving $P$.
In the sense of polynomial complexity, in such a case $Q$ is at least as hard to solve as $P$.
If the same holds for $Q$ and any other problem instead of $P$ in \NP,
then $Q$ is the hardest of all problems in \NP, in the above sense. Such a problem $Q$ in \NP\ is an \NP-complete problem.

\citet{Cook1971} and  
\citet{Levin1973} independently showed that there are \NP-complete problems. They each proved that
the unrestricted Boolean satisfiability problem (SAT) is \NP-complete.
This was a major breakthrough, because it allowed many other problems to be shown to be \NP-complete, by using a
 polynomial reduction from a known \NP-complete problem (starting with SAT)
to the newly considered problem in \NP\ \citep{Karp1972}. 
The TSP and $k$-SAT (with $k\ge 3$) decision problems (section~\ref{sec:decision}) are both \NP-complete; the 2-SAT problem is in \Poly.

If a polynomial algorithm exists for any of these \NP-complete problems, then a polynomial algorithm would exist for each of them, by using the reduction process used in their proofs of \NP-completeness.
The existence of an ever growing number of \NP-complete problems
for which nobody to date has been able
to develop a polynomial algorithm provides significant evidence (although not proof) supporting the conjecture {\Poly} $\neq$ \NP.

\subsection{\NP-hard problems}
For problems other than decision problems, such as the optimisation and counting problems mentioned earlier,
their computational complexity is usually defined only if they are in \NP\ and contain decision problems in \NP\ as special cases, and hence are at
least as difficult to solve as their decision counterparts. 
Such problems are called \NP-\textit{hard} if they are at least as hard
as an \NP-complete problem, that is, if a polynomial time algorithm for solving them would imply a
polynomial algorithm for solving an \NP-complete (decision) problem.
For a compendium of \NP-hard optimisation problems, see \citet{Nada_KTH}. 

\subsection{Other classic complexity classes: \PSPACE\ and \BPP}

Since the introduction of the \Poly\ and \NP\ complexity classes, a whole zoo of further complexity classes has been defined and studied. Most of these classes are beyond the scope of this chapter, but we mention a few here that are relevant in the context of this chapter.

The complexity class
\PSPACE\ consists of decisions problems that can be solved using polynomial space  on a DTM, meaning that the number of cells on the tape of the DTM that are needed to encode and solve a given instance is bounded by a polynomial function in the length of the input size.  Note that no constraint is put on the time allowed for the solution (other than being finite). For this class, using an NTM does not add any extra computational power in terms of space use, because an NTM that uses polynomial space can be simulated by a DTM that uses (more but still) polynomial space (but it may use substantially more time).

We clearly have {\NP} $\subseteq$ \PSPACE: if a problem can be checked in polynomial time, it cannot use more than polynomial space, since it has to visit all of that space.
It is widely believed that {\NP} $\neq$ \PSPACE, but again, there is no proof.

The complexity class
\BPP\ (Bounded-error Probabilistic Polynomial) consists of decision problems that can be solved in polynomial time by a \textit{probabilistic} Turing Machine (PTM), i.e., a DTM that can make random choices between different transitions according to some probability distribution. 
(This is distinct from an NTM:
a probabilistic TM makes a \textit{random} choice; an NTM makes the `correct' choice, or all choices, depending on the interpretation.)
The probability that any run of the algorithm gives the wrong answer to a {\sc yes-no} question must be less than $1/3$.

It is obvious that {\Poly} $\subseteq$ \BPP: if a problem can be solved in polynomial time, it can be probabilistically solved in polynomial time.
In this case, it is widely believed that {\Poly} $=$ \BPP, but yet again, there is no proof.
There is no known subset relation between \BPP\ and \NP, in either direction.

\subsection{Quantum complexity classes}
\label{sec:BPQ}

A \textit{quantum} TM (QTM), with a quantum processor and quantum tape (memory) is a model for a quantum computer, computing directly in memory \citep{Deutsch1985}.

\begin{figure}[tp]
\includegraphics[width=0.5\linewidth]{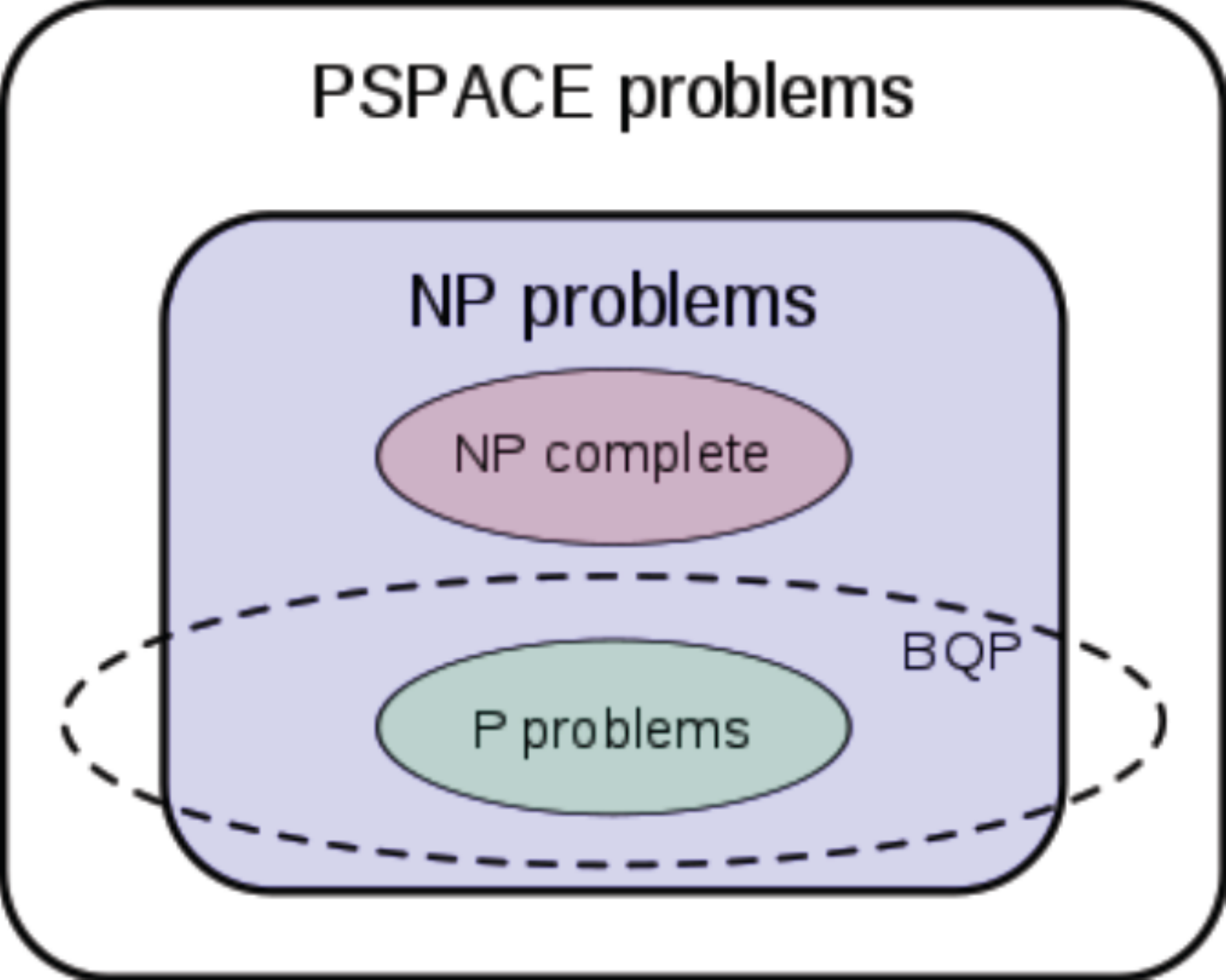}
\centering
\caption{
Summary of relationships between computational complexity classes (https://en.wikipedia.org/wiki/BQP). See text for details.
}
\label{BQP}
\end{figure}

Problems that can be 
solved by a QTM in polynomial time belong to the complexity class  \BQP\ (Fig.~\ref{BQP}) \citep{CplxZoo2014,Watrous2009,Montanaro2016}. \BQP\ is in some sense the quantum analogue of the classical  \BPP, but there is a fundamental difference: the PTM proceeds via random choices of unique states of the Finite State Machine reading and writing on the tape, while the QTM proceeds via quantum simultaneous superposition and entanglement of all the states. Therefore, the QTM proceeds through the Hilbert state space in a {\em deterministic} way via the time evolution operator defined by the Hamiltonian. The probabilistic aspects emerge when reading out the results; in some cases this can be done deterministically, in other cases one has to collect statistics.

Fig.~\ref{BQP} shows that \BQP\ is a limited region of the complexity map,  and (probably) does not include the \NP-complete class. 
There, in \BQP\ and (probably) outside \Poly, we find problems like Shor's factorisation algorithm \citep{Shor1994}, providing exponential speed-up over the best known classical factorisation algorithm.
Classic factorisation is believed to be neither \NP-complete, nor in \Poly. Another example is unstructured database search, which is classically ``easy" (polynomial), but which shows quadratic speed-up with the Grover quantum search algorithm \citep{Grover1996}. See  \citet{Montanaro2016} for a recent  overview of progress in the field, focusing on algorithms with clear applications and rigorous performance bounds.

There are further quantum complexity classes.  In particular, \QMA\ is the class where problems proposed by a quantum oracle can be verified in polynomial time by a quantum computer in \BQP\ \citep{CplxZoo2014,Watrous2009,Montanaro2016}.

\subsection{Complexity with advice: \Poly/poly and \Poly/log}

The most common complexity class involving advice is \Poly/poly, where the advice length $f(n)$ can be any polynomial in $n$. This class \Poly/poly is equal to the class consisting of decision problems for which there exists a polynomial size Boolean circuit correctly deciding the problem for all inputs of length $n$, for every $n$. 
This is true because a DTM can be designed that interprets the advice string as a description of the Boolean circuit, and conversely, a (polynomial) DTM can be simulated by a (polynomial) Boolean circuit.

Interestingly, \Poly/poly contains both \Poly\ and \BPP\, and it also contains some undecidable problems (including the \textit{unary} version of the Halting Problem). It is widely believed that \NP\ is not contained in \Poly/poly, but again there is no proof for this.
If has been shown that \NP\ $\not\subset$ \Poly/poly implies \Poly\  $\neq$ \NP. Much of the efforts towards proving that \Poly\ $\neq$ \NP\ are based on this implication.

The class \Poly/log is similar to \Poly/poly, except that the advice string for inputs of size $n$ is restricted to have length at most logarithmic in $n$, rather than polynomial in $n$.
It is known that \NP\ $\subseteq$ \Poly/log implies \Poly\ = \NP.

Restricting the advice length to at most a logarithmic function of the input size implies that polynomial reductions cannot be used to show that a decision problem belongs to the class \Poly/log. To circumvent this drawback the prefix advice class Full-\Poly/log has been introduced \citep{Balcazar1998}. 
The difference with \Poly/log is that in Full-\Poly/log each advice string for inputs of size $n$ can also be used for inputs of a smaller size. Full-\Poly/log is also known as \Poly/log* in the literature.

\subsection{Extended Church-Turing thesis}

The CTT (section~\ref{sec:CTT}) is a statement about \textit{computability}.
The {\em Extended Church-Turing Thesis} (ECT) is a statement about \textit{complexity}: any function naturally to be regarded as {\em efficiently} computable is efficiently computable by a DTM  \citep{Dershowitz2012}. Here ``efficiently'' means computable by a DTM in polynomial  time and space. A DTM is a basic model for ordinary {\em classical} digital computers solving problems tractable in polynomial time. 

Consider the NTM (section~\ref{sec:PandNP}).  If backtracking is the most efficient way to implement an NTM with a DTM (that is, if \Poly\ $\neq$ \NP), then the ECT claims that a `true' NTM cannot be implemented.

\citet{Vergis1986} reformulate the ECT in terms of analogue computers as:
``any finite analogue computer can be simulated {\em efficiently}  by a digital
computer, in the sense that the time required by the digital computer to simulate the analogue computer
is bounded by a polynomial function of the resources used by the analogue computer''.
That is, finite analogue computers do not make infeasible problems feasible.
 Thus,
finite analogue computers cannot tractably solve \NP-complete problems.

Super-Turing computation seeks to discover approaches that can expand the range of efficiently computable functions beyond those efficiently computable by DTMs; it seeks to invalidate the ECT.

Quantum computing (QCOMP) can provide exponential speed-up for a few classes of problems (see \S\ref{sec:BPQ}),
so the ECT is believed to have been invalidated in this case \citep{Aaronson2013,Aaronson2013a}: quantum computing can provide certain classes of super-Turing power (unless \Poly\ $=$ \NP).  We can extend the ECT to: ``any function naturally to be regarded as {\em efficiently} computable is efficiently computable by a Quantum Turing Machine (QTM)'', and then ask if any \textit{other} form of UCOMP can invalidate either the original ECT, or this quantum form of the ECT.  

\subsection{Physical oracles and advice}
\label{sec:phys-oracle}

An interesting question in UCOMP is whether a \textit{physical} system can be implemented that acts as some specific oracle or advice, that is, whether it is possible to build a physical add-on to a classical computer that can change the computability or complexity classes of problems.

A range of analogue devices have been posited as potential physical oracles.
Their analogue physical values may be (theoretically) read with infinite, unbounded, or fixed precision, resulting in different (theoretical) oracular power.

Beggs and coauthors (see \citep{Ambaram-2017} and references therein) have made a careful analysis of using a range of idealised physical experiments as oracles,
in particular, studying the time it takes to interact with the physical device, as a function of the precision of its output.  More precision takes more time.
In each system they analyse, they find that the time
needed to extract the measured value of the analogue system increases exponentially with the number of bits of precision of the
measurement.  They conjecture that
\begin{quote}
for all ``reasonable'' physical theories and for all measurements based on
them, the physical time of the experiment is at least exponential, i.e., the time needed
to access the $n$-th bit of the parameter being measured is at least exponential in $n$.
\end{quote}

\begin{figure}[tp]
\centering
\includegraphics[width=0.98\linewidth,trim = 27mm 107mm 66mm 72mm, clip]{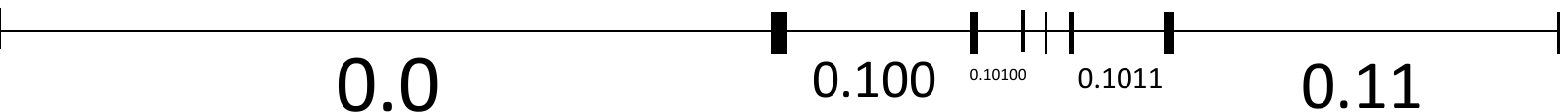}
\caption{\label{fig:unary}Encoding a real number value as a position on a line.
Extraction of each extra digit of precision requires access to an exponentially smaller region of the line.
}
\end{figure}

The kind of physical analogue devices that \citet{Ambaram-2017} analyse tend to use a \textit{unary} encoding of the relevant value being accessed via physical measurement, for example, position, or mass, or concentration.
So each extra digit of precision has to access an exponentially smaller range of the system being measured. See fig.~\ref{fig:unary}.

Similar points apply when discussing the input size $n$ when analysing such devices: if the UCOMP device uses a unary encoding of the relevant parameter values, as many do, the input size $n$ is exponentially larger than if a binary encoding were used.

\citet{Beggs2014} use such arguments to derive an \textit{upper bound} on the power of such hybrid analogue-digital machines, and conjecture an associated ``Analogue-digital Church-Turing Thesis'':
\begin{quote}
No possible abstract analogue-digital device can have more computational capabilities in polynomial time
than \BPP//log*.
\end{quote}
Note that this conjecture refers to abstract (or idealised) physical devices, analogous to the way a DTM is also an abstract idealised device. Physical issues such as thermodynamic jitter and quantum uncertainty have still to be considered.
Note also that the logarithmic advice has to be encoded somehow into the analogue device.

\citet{Blakey2014,Blakey2017} has made a careful analysis of ``unconventional resources'' such as precision (measurement and manufacturing), energy, and construction material, and has developed a form of ``model-independent'' complexity analysis, that considers such unconventional resources in addition to the classical ones of space and time. 
Indeed, Blakey's analysis shows that the exponential cost of measurement precision is itself outweighed by the \textit{infinite} manufacturing precision required for certain devices.
Blakey's approach to analysing unconventional resources enables a formalisation of ``the intuition that the purported [hypercomputational] power of these computers in fact
vanishes once precision is properly considered''.

\subsection{Complexity as a worst case property}
\label{sec:worst}

\NP-completeness is a worst case analysis: a problem is \NP-complete if it has at least \textit{one} instance that requires exponential, rather than polynomial, computation time, even if all the remaining instances can be solved in polynomial time.

\citet{Cheeseman1991} note that for many \NP-complete problems, typical cases are often easy to solve, and hard cases are rare.  They show that such problems have an ``order parameter'', and that the hard problems occur at the critical value of this parameter.
Consider 3-SAT (section~\ref{sec:decision}); the order parameter is the average number of constraints (clauses) per Boolean literal, $m/n$.
For low values the problem is underconstrained (not many clauses compared to literals, so easily shown to be satisfiable) and for high values it is overconstrained (many clauses compared to literals, so easily shown to be unsatisfiable).  Only near a critical value do the problems become exponentially hard to determine. 

Such arguments demonstrate that we cannot `sample' the problem space to demonstrate problem hardness; complexity is not an experimental property.
In particular, demonstrating that a device or process, engineered or natural, can solve some (or even many) \NP-complete problem instances tractably is not sufficient to conclude that it can solve all \NP-complete problem instances tractably.

The limitations that \NP-completeness imposes on computation probably hold for all natural analogue systems, such as protein folding, the human brain, etc. \citep{Bryngelson1995}. As noted above, just because Nature can efficiently solve \textit{some} instances of problems that are \NP-complete does not mean that it can solve \textit{all} \NP-complete problem instances \citep{Bryngelson1995}.
To find the lowest free energy state of a general macromolecule has been shown to be \NP-complete \citep{Unger1993}. In the case of proteins there are amino acid sequences that cannot be folded to their global free energy minimum in polynomial time either by computers or by Nature. Proteins selected by Nature and evolution will represent a \textit{tractable subset} of all possible amino acid sequences.

\subsection{Solving hard problems in practice}

Apart from trying to show that {\Poly} $=$ \NP, there are other seemingly more practical ways to try to cope with
\NP-complete or \NP-hard problems.

If large instances have to be solved, one approach is to look for fast algorithms, called \textit{heuristics}, that
give reasonable solutions in many cases. In some cases there are approximation algorithms for optimisation problems
with provable approximation guarantees. This holds for the optimisation variant of the TSP restricted to instances for which the triangle equality holds (the weight of edge $uv$ is at most the sum of the weights of the edges $uw$ and $wv$, for all distinct triples of vertices $u,v,w$), and where one asks for (the length of) a shortest tour. This variant is known to be \NP-hard, but simple polynomial time heuristics have been developed that yield solutions within a factor of 1.5 of the optimal tour length \citep{Lawler1985}.

For many optimisation problems even guaranteeing certain approximation bounds is an \NP-hard problem in itself. This also holds for the general TSP (without the triangle inequality constraints) if one wants to find a solution within a fixed constant factor of the optimal tour length \citep{Lawler1985}.

A more recent approach tries to capture the exponential growth of solution algorithms in terms of
a function of a certain fixed parameter that is not the size of the input. The aim is to develop a solution algorithm
that is polynomial in the size of the input but maybe exponential in the other parameter. For small values of the
fixed parameter the problem instances are tractable, hence the term fixed parameter tractability (the class \FPT) for such problems \citep{Downey1999}.

An example is $k$-SAT (section~\ref{sec:decision}), parameterised by the number $n$ of Boolean literals. A given formula of size $N$ with $n$ literals can be checked by brute force in time $O(2^{n}N)$, so linear in the size of the instance.

A related concept is that of preprocessing (data reduction or
kernelisation). Preprocessing in this context means reducing the input size of the problem instances to something 
smaller, usually by applying reduction rules that take care of easy parts of the instances. Within parameterised complexity theory,
the smaller inputs are referred to as the kernel.
The goal is to prove that small kernels for certain \NP-complete or
\NP-hard problems exist, and can be found in polynomial time. If small here means bounded by a function that only depends on some 
fixed parameter associated with the
problem, then this implies that the problem is fixed parameter tractable.

The above definitions focus on worst case instances of the (decision) problems.  
It is not clear whether this is always a practical focus. 
There is a famous example of a class of problems in \Poly\ -- Linear Programming -- for which empirical evidence shows that
an exponential algorithm (the Simplex Method) for solving these problems very often yields faster solutions in practice than the polynomial algorithm (the Ellipsoid Method) 
 developed subsequently \citep{Papadimitriou1994}. 

For many algorithms, a worst case analysis gives limited insight into their performance,  and can be far too pessimistic to reflect the actual performance on realistic instances. 
Recent approaches to develop a more realistic and robust model for the analysis of the performance of algorithms include average case analysis,
smoothed analysis, and semi-random input models. All of these approaches are based on considering instances that are to a certain extent randomly chosen.

\subsection{No Free Lunch theorem}

\citet{Wolpert1997} prove ``no free lunch'' (NFL) theorems related to the efficiency of search algorithms.
They show that when the performance of any given search algorithm is averaged over all possible search landscapes, it performs no better than random search.
This is because, whatever algorithm is chosen, if it exploits the structure of the landscape, there are always deceptive search landscapes that lead it astray.
The only way not to be deceived is to search randomly.
A problem of size $n$ has a search space of size $O(2^n)$, and so the best classical search algorithm, where the performance is averaged over all the $O(2^{2^n})$ possible landscapes, is $O(2^n)$.

This does not mean that there are no search algorithms better than random search over certain subsets of search landscapes:
algorithms that exploit any structure common across the subset can perform better than random.  
More generally, if some search landscapes are more likely than others, algorithms that can exploit that information can do better \citep{Wolpert2012}.

Natural processes such as Darwinian evolution, which may be interpreted as a form of search algorithm, are almost certainly exploiting the structure of their search landscapes.
This has consequences for nature-inspired search algorithms, such as evolutionary algorithms, if they are to be exploited on `unnatural' landscapes.  See also the comments on protein folding in section~\ref{sec:worst}.

\section{Quantum information processing}
\label{sec:qip}

Quantum computers are able to solve some problems much faster than classical computers \citep{CplxZoo2014,Shor1994,Watrous2009,Montanaro2016}. However, this does not say much about  solving computational problems that are hard for classical computers. If one looks at the map of computational complexity (Fig.~\ref{BQP}), classifying the hardness of computational (decision) problems, one finds that the \BQP\ class of quantum computation covers a rather limited space, not containing really hard problems. One may then ask what is the fundamental difference between CCOMP and QCOMP, and what kind of problems are hard even for a quantum computer? \citep{Aaronson2005,Aaronson2008,Aaronson2009,Aaronson2013}

\subsection{Digital quantum  computation}

\begin{figure}[tp]
\includegraphics[width=0.95\textwidth]{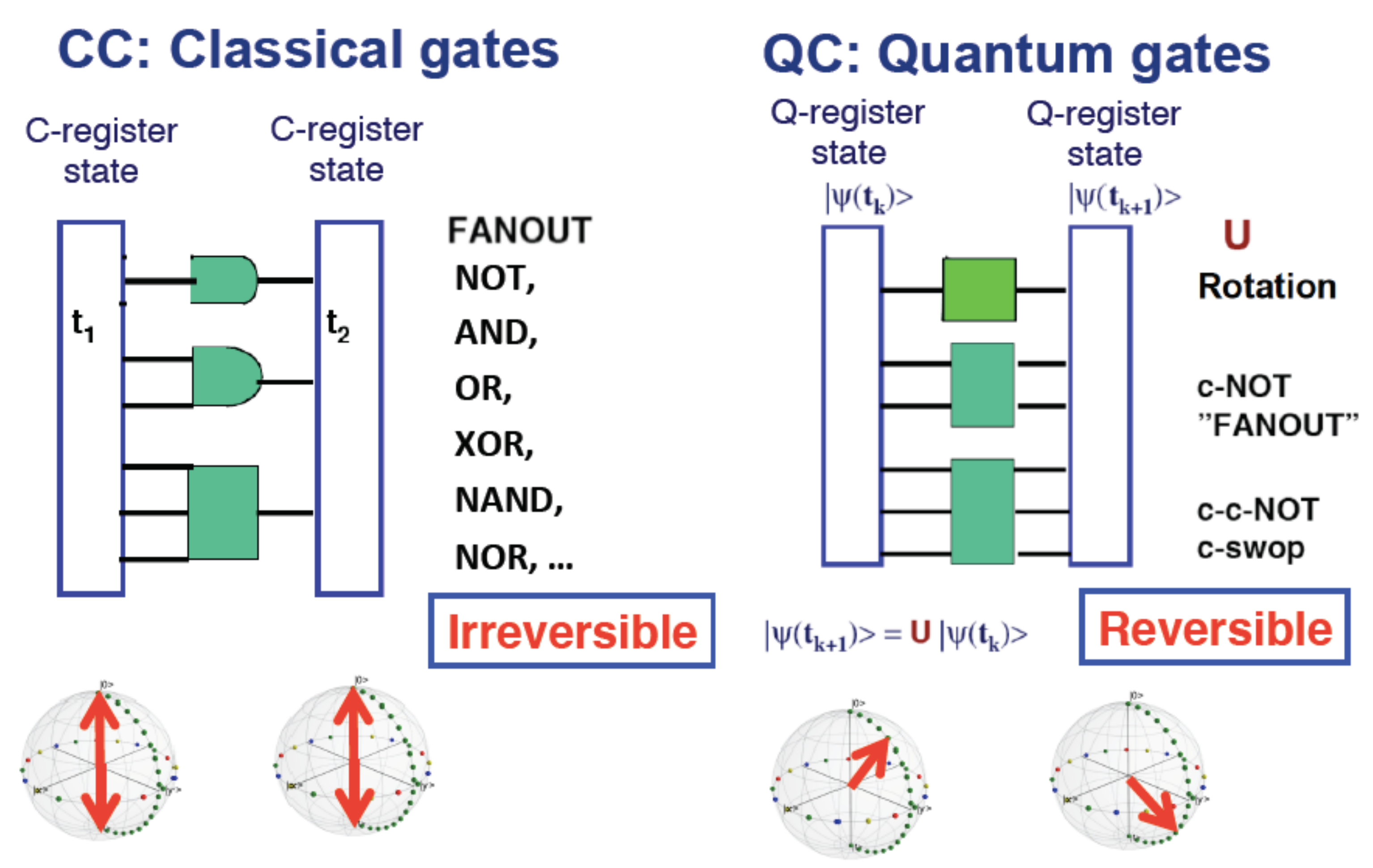}
\centering
\caption{{\bf a.}
Comparison of CCOMP and QCOMP. (left) CCOMP: irreversible gates with arithmetic-logic unit (ALU) and memory separated. The memory is the storage, with classical bits 0,1 representing the poles on the Bloch unit sphere. Classical gates are basically hardwired, irreversible and performed in the ALU.  Gates are clocked. 
(right) QCOMP: Computing in memory -- the memory is the computer.  Quantum bits (qubits)  $\alpha \ket{0}+\beta \ket{1}$ span the entire Bloch sphere. Quantum gates are reversible and performed on the ``memory'' qubits by software-controlled external devices. Gates are not clocked. 
}
\label{CCQC}
\end{figure}

The obvious difference between CCOMP and QCOMP is that CCOMP is based on classical Newtonian physics and special and general relativity, while QCOMP is based on quantum physics, as illustrated in Figs.~\ref{CCQC}a,b. 
Digital CCOMP progresses by gate-driven transitions between specific classical memory configurations of an N-bit register $R(t_k)$, each representing one out of $2^N$ instantaneous configurations. 

\addtocounter{figure}{-1}
\begin{figure}[tp]
\includegraphics[width=0.95\textwidth]{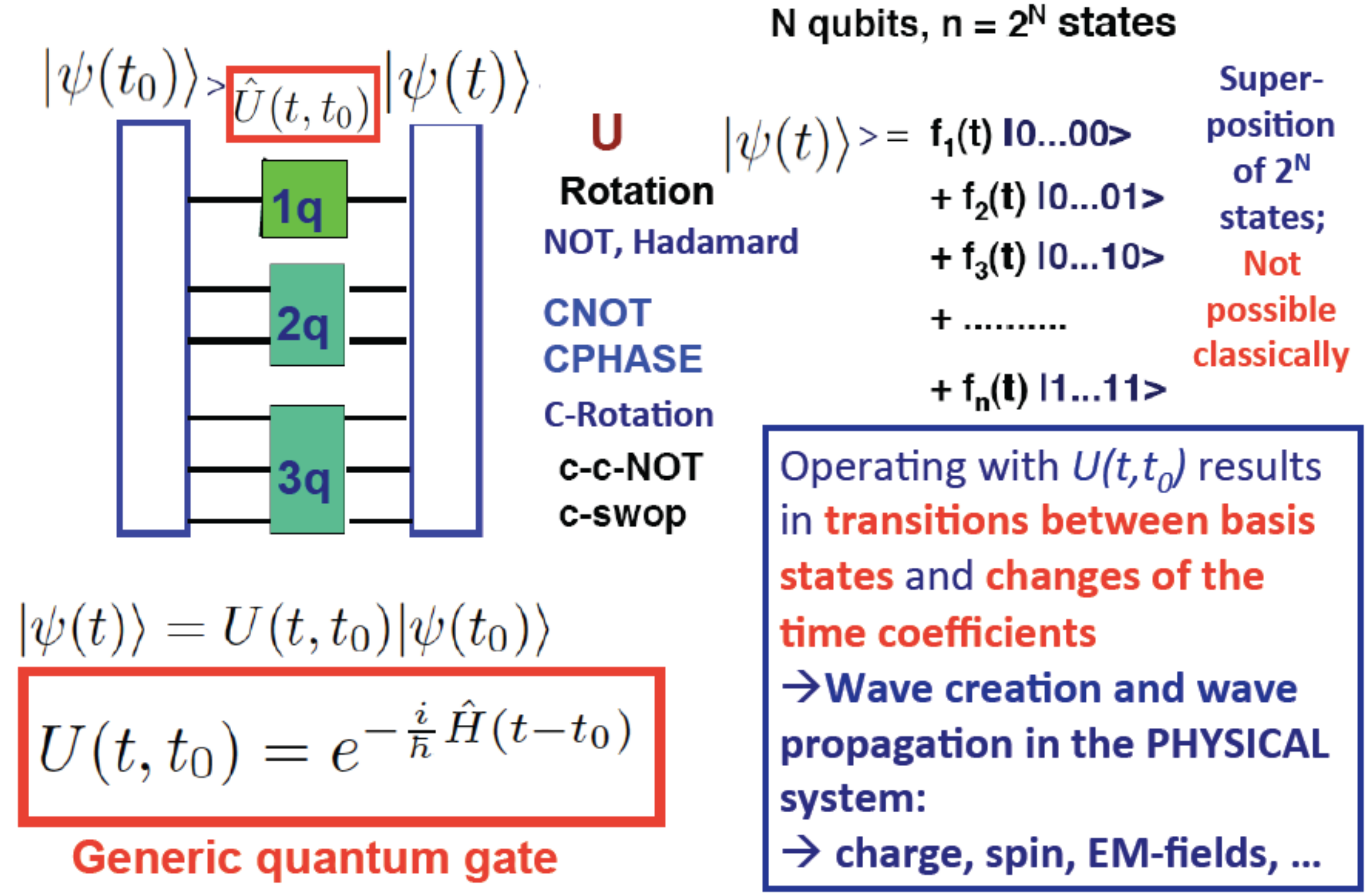}
\centering
\caption{{\bf b.}
Comparison of CCOMP and QCOMP (ctd). 
(left) The time evolution of the state of the quantum computer is implemented by the unitary time evolution operator U, which can be broken down into elementary gates (like NOT, Hadamard, CNOT, C-Rotation). 
(right) The quantum state is, in general, an entangled superposition of all configurations of the memory register. Entanglement implies that the state is not a product state, containing non-classical correlations that provide polytime solution of certain problems that take exponential time for classical computers. The quantum gates are operated by a classical computer, which means that a quantum computer can only solve problems that take at most a polynomial number of gates to solve. A universal set of quantum gates (Hadamard (square-root of bit flip), X (bit flip), CNOT, and T (general rotation)) guarantees the existence of a universal QCOMP, like a UTM, but may likewise need exponential resources.
}
\label{CCQC-ctd}
\end{figure}

QCOMP, on the other hand, progresses by gate-driven transitions between specific quantum memory states $\ket{\Psi(t_k)}$, each representing instantaneous superposition of $2^N$  configurations. The quantum memory states are coherent amplitudes with well-defined phase relations. Moreover, the states of the qubits can be entangled, i.e., not possible to write as a product of states.
In the case of two qubits, the canonical example of entanglement is that of Bell states: non-entangled product states are $\ket{00}$ or $\ket{11}$ or $(\ket{0}+\ket{1})(\ket{0}+\ket{1})$ = $\ket{00}+\ket{01}+\ket{10}+\ket{11}$. The Bell states $\ket{00}\pm\ket{11}$ and $\ket{0}\pm\ket{10}$ are clearly not product states, and represent in fact maximum entanglement of two qubits. This can be generalised to more qubits, e.g., the Greenberger-Horne-Zeilinger (GHZ) "cat state": $\ket{000}+\ket{111}$. This entanglement represents non-classical correlations, at the heart of the exponential power of QCOMP.
Entanglement allows us to construct maximally entangled superpositions with only a linear amount of physical resources, e.g., a large cat state: $\frac{1}{\sqrt{2}} (\ket{0......00} +  \ket{1.....11})$. This is what allows us to perform non-classical tasks and provide quantum speed-up \citep{JozsaLinden2003,Horodecki2009}. 

QCOMP is basically performed directly in memory. One can regard the qubit memory register as an array of 2-level quantum transistors, memory cells,  where the gates driving and coupling the transistors  are external {\em classical} fields controlled by classical software run on a CCOMP. This emphasises that a quantum computer can only implement a polynomial number of gates, and that the name of the game is to devise {\em efficient} decompositions of  the time-evolution operator in terms of universal gates. The goal is of course to construct single-shot multi-qubit gates implementing long sequences of canonical elementary gates to synthesise the full time-evolution operator $U = exp(-iHt)$ (Fig.~\ref{CCQC}). 

QCOMP depends, just like CCOMP, on encoding/decoding, error correction, and precision of measurement and control. To go beyond \BQP\ essentially takes non-physical oracle resources, or unlimited precision, requiring exponential resources and ending up in \QMA\ or beyond. 

\subsection{Quantum simulation} 
\citet{Feynman1982} was among the first to point out that quantum systems need to be described by quantum systems. Electronic structure calculation with full account of many-body interactions is \QMA-hard \citep{Aaronson2009,Schuch2009,Whitfield2013}.
Therefore, taking an example from biochemistry, to efficiently compute the properties of a catalysing enzyme or the workings of a ribosome will require a quantum simulator to achieve the precision needed in reasonable time. 

A QCOMP emulator/simulator is basically an analogue machine: an engineered qubit-array where the interactions between qubits (memory cells) are implemented by substrate-defined or externally induced local and global couplings. The static or quasi-static (adiabatic)  interactions can be tuned to implement specific Hamiltonians describing physical models or systems  (or even something unphysical), and time-dependent driving will implement dynamic response. All the interactions provide together an effective time-dependent Hamiltonian and a corresponding  time-evolution operator $U = exp[-iH_{eff}(t)\ t\ ]$. The induced time-evolution will be characteristic for the system and can be analysed (measured) to provide deterministic or statistical answers to various questions. Note that there is no fundamental difference between digital and analogue QCOMP: if we drive the qubits by, e.g., fast external microwave pulse trains, we can design the time-evolution operator $U$ to generate the specific elementary universal 1q and 2q gates of the quantum circuit model. 

Quantum simulation of physical systems \citep{Brown2010,Georgescu2014,Wendin2016} is now at the focus of intense engineering and experimental efforts 
\citep{CiracZoller2010,Lanyon2010,Barreiro2011,BlattRoos2012,Peruzzo2014,Wang2015,Salathe2015,Barends2015,Barends2016,OMalley2016,Boixo2016} and software development \citep{Wecker2014,Valiron2015,Haner2016a,Haner2016b,Bauer2016,Reiher2016,BravyiGosset2016}. Materials science and chemistry will present testing grounds for the performance of quantum simulation and computing in the coming years. 

\subsection{Adiabatic Quantum Optimisation (AQO)}
AQO is the analogue version of quantum computing and simulation. It starts from the ground state of a simple known Hamiltonian and slowly (adiabatically) changes the substrate parameters into describing a target Hamiltonian, manoeuvring through the energy landscape, all the time staying in the ground state. The final state and the global minimum then present the solution to the target problem \citep{Farhi2014,FarhiHarrow2016}. AQO is potentially an efficient approach to quantum simulation but has so far been limited to theoretical investigations, e.g.,  with applications to quantum phase transitions and speed limits for computing.  Possibly there is a Quantum No-Free-Lunch theorem stating that digital quantum gate circuits need quantum error correction and AQO needs to manoeuvre adiabatically through a complicated energy landscape, and in the end the computational power is the same. 

\subsection{Quantum annealing (QA)}
QA is a version of quantum optimisation where the target Hamiltonian (often a transverse Ising Hamiltonian) is approached while simultaneously lowering the temperature. This is the 
scheme upon which D-Wave Systems have developed their QA processors, the most recent one built on a chip with a 2000 qubit array and a special cross-bar structure. Despite indications of quantum entanglement and tunnelling within qubit clusters \citep{Denchev2015,Boixo2016a}, there is no evidence for quantum speed-up \citep{Ronnow2014,Zintchenko2015} -- so far optimised classical algorithms running on modest classical computers can simulate the quantum annealer.

\subsection{Quantum machine learning (QML)} 
QML is an emerging field, introducing adaptive methods from machine language (ML) classical optimisation and neural networks to quantum networks \citep{Schuld2015,Wiebe2014,Wiebe2015a,Aaronson2015,Wittek2016,Biamonte2016}. One aspect is using ML for optimising classical control of quantum systems. Another, revolutionary, aspect is to apply ML methods to quantum networks for quantum enhanced learning algorithms. The field is rapidly evolving, and we refer to a recent review \citep{Biamonte2016} for an overview of progress and for references.

\section{Computational power of classical physical systems and unconventional paradigms}
\label{sec:power}

As already mentioned in the Introduction, and discussed to some extent, there is a veritable zoo of UCOMP paradigms \citep{UCOMP2009}. Here we claim that the only decisive borderline is the one that separates classical problems (Newtonian Physics and Relativity) from problems governed by Quantum Physics, which includes some combinatorial problems profiting from the Quantum Fourier Transform (QFT). Quantum information processing and class \BQP\ is discussed in \S\ref{sec:qip}. In this section we focus on a few classical problems of great current interest, representative for the  polynomial class \Poly.

There are further issues of measuring the complexity of problems running on UCOMP devices.
The actions of the device might not map well to the parameters of time, space, and problem size needed for classical complexity analysis.
And the computation may use resources not considered in classical complexity analysis, for example, the time needed to read out a result.

\subsection{DNA computing}

Computing with DNA or RNA strands was first investigated theoretically. \citet{Bennett1982} imagines a DTM built from RNA reactions:
\begin{quote}
The tape might be a linear informational macromolecule analogous to RNA, with an additional chemical group attached at one site to encode the head state \ldots and location.  Several hypothetical enzymes (one for each of the Turing Machine's transition rules) would catalyse reactions of the macromolecule with small molecules in the surrounding solution, transforming the macromolecule into its logical successor.
\end{quote}
\citet{Shapiro2012} proposes a more detailed design for a general purpose polymer-based DTM.
\citet{Qian2010} describe a DNA-based design for a stack machine.
These designs demonstrate that general purpose polymer-based computing is possible, at least in principle.  None of these designs challenge the ECT: they are all for DTMs or equivalent power machines.

\citet{Adleman1994} was the first to implement a form of DNA computing in the wetlab,
with his seminal paper describing the solution to a 7-node instance of the Hamiltonian path problem.
This is an \NP-complete decision problem on a graph: is there a path through a graph that visits each node exactly once?
Adleman's approach encodes the graph nodes and edges using small single strands of DNA, designed so that the edges can stick to the corresponding vertices by complementary matching.
Sufficient strands are put into a well mixed system, and allowed to stick together.  A series of chemical processes are used to extract the resulting DNA, and to search for a piece that has encoded a solution to the problem.  The time taken by these processes is linear in the number of nodes,
but the number of strands needed to ensure the relevant ones meet and stick with high enough probability grows exponentially with the number of nodes \citep{Adleman1994}.
Essentially, this set-up needs enough DNA to construct all possible paths, to ensure that the desired solution path is constructed.
So this algorithm solves the problem in polynomial time, by using massive parallelism, at the cost of exponential DNA resources (and hence exponential space).
\citet{Hartmanis1995} calculates that this form of computation of the Hamiltonian path on a graph with 200 nodes would need a mass of DNA greater than that of the earth.

\citet{Lipton1995} critiques Adleman's algorithm, because it is ``brute force'' in trying all possible paths.  He describes a (theoretical) DNA algorithm to solve the \NP-complete SAT problem (section~\ref{sec:decision}).  For a problem of $n$ Boolean variables and $m$ clauses, the algorithm requires a number of chemical processing steps linear in $m$.
However, it also requires enough DNA to encode ``all possible $n$-bit numbers'' (that is, all possible assignments of the $n$ Boolean variables), so it also requires exponential DNA resources. 

So these special-purpose forms of DNA computing, focussed on \NP-complete problems, trade off exponential time for exponential space (massively parallel use of DNA resources).
Additionally, it is by no means clear that the chemical processing and other engineering facilities would remain polynomial in time once the exponential physical size of the reactants kicks into effect.

Other authors consider using the informational and constructive properties of DNA for a wide range of computational purposes.
For example, implementations of tiling models often use DNA to construct the tiles and program their connections.

The Wang tile model \citep{Wang1961}
has been used to show theorem proving and computational capabilities within, e.g., DNA computing. It was introduced by Wang, who posed several conjectures and problems related to the question whether a given finite set of Wang tiles can tile the plane. 
Wang's student Robert Berger 
 showed 
how to emulate any DTM by a finite set of Wang tiles \citep{Berger1966}. Using this, he proved that the undecidability of the Halting Problem implies the undecidability of Wang's tiling problem.

Later applications demonstrate the
(computational) power of tiles; see, e.g., 
\citet[ch.6]{Yang2011}. In particular, \NP-complete problems like $k$-SAT have been solved in linear time in the size of the input using a finite number of different tiles \citep{Brun2008}. 
The hidden complexity lies in the exponentially many parallel tile assemblies (the computation is nondeterministic and each parallel assembly executes in time linear in the input size).

As for the latter example, in each of these models the complexity properties need to be carefully established.
Properties established in one implementation approach may not carry over to a different implementation.
For example, \citet{Seelig2011} consider
molecular logic circuits with many components arranged in multiple layers built using DNA strand displacement;
they show that the time-complexity does not necessarily scale linearly with the circuit depth, but rather can be quadratic, and that catalysis can alter the asymptotic time-complexity.

\subsection{Networks of Evolutionary Processors}

An ``Accepting Hybrid Network of Evolutionary Processors'' (AHNEP) is a theoretical device for exploring language-accepting processes. \citet{Castellanos2001} describe the design.
It comprises a fully connected graph.
Each graph node contains (i) a simple evolutionary processor that can perform certain point mutations (insertion, deletion, substitution) on data, expressed as rewrite rules; (ii) data in the form of a multiset of strings, which are processed in parallel such that all possible mutations that can take place do so.  In particular, if a specified substitution may act on different occurrences of a symbol in a string, each occurrence is substituted in a different copy of the string.  For this to be possible, there is an arbitrarily large number of copies of each string in the multiset.
The data moves through the network; it must pass a filtering process that depends on conditions of both the sender and receiver. 

\citet{Castellanos2001} demonstrate that such networks of linear size (number of nodes) can solve \NP-complete problems in linear time.  Much subsequent work has gone into variants \citep{Margenstern2005}, and determining bounds on the size of  such networks.
For example, \citet{Manea2007} find a constant size network of 24 nodes that can solve \NP\ problems in polynomial time; \citet{Loos2009} reduce that bound to 16 nodes. 
\citet{Alhazov2014} prove that a 5 node AHNEP is computationally complete.

Note that some of the papers referenced demonstrate that AHNEPs can solve \NP-complete problems in polynomial time.
They manage to do so in the same way the DNA computers of the previous section do:
by exploiting exponential space resources.
The set-up exploits use of an exponentially large data set at each node, by requiring an arbitrarily large number of each string be present, so that all possible substitutions can occur in parallel.

\subsection{Evolution in materio}

Evolution \textit{in materio} (EiM) is a term coined by \citet{Miller2002} to refer to material systems that can be used for computation by manipulating the state of the material through external stimuli, e.g., voltages, currents, optical signals and the like, and using some fixed input and output channels to the material for defining the wanted functionality. In EiM, the material is treated as a black box, and computer-controlled evolution (by applying genetic algorithms or other optimisation techniques, using a digital computer) is used to change the external stimuli (the configuration signals) in such a way that the black box converges to the target functionality, i.e., the material system produces the correct output combinations, representing solutions to the problem
  when certain input combinations, representing problem instances of the problem,
  are applied. Experimental results show that this approach has successfully been applied to different types of problems, with different types of materials \citep{Broersma2016}, but mainly for either small instances of problems or for rather simple functionalities like Boolean logic gates. Interestingly, using EiM, reconfigurable logic has been evolved in a stable and reproducible way on disordered nanoparticle networks of very small size, comparable to the size that would be required by arrangements of current transistors to show the same functionality \citep{Bose2015}.
  
It is impossible and it would not be fair to compare the complexity of the solution concept of EiM to that of classical computation.
First of all, apart from the (digital)
genetic algorithms or other optimisation techniques that are used to manipulate the system, there is no algorithmic procedure involved in the actual computation. The material is not executing a program to solve any particular problem instance; instead, a set of problem instances and their target outputs are used in the evolutionary process of configuring the material.
In that sense, the material is more or less forced to produce the correct (or an approximate solution that can be translated into a correct) solution for that set of problem instances. If this is not the whole set of possible instances, there is no guarantee that the material outputs the correct solution for any of the other problem instances. In fact, since fitness functions are used to judge the quality of the configurations according to the input-output combinations they produce, even the correctness of the output for individual instances that are used during the evolutionary process is questionable, unless they are checked one by one at the end of this process. In a way, for problems with an unbounded number of possible instances, the EiM approach can be regarded as a heuristic without any performance guarantees for the general problem. So, it is not an alternative to exact solution concepts from classical computation, and hence it cannot claim any particular relevance for (exactly) solving \NP-hard or \NP-complete problems, let alone undecidable problems.

Secondly, in EiM the time it takes for the evolutionary process to converge to a satisfactory configuration of the material for solving a particular problem is the crucial measure in terms of time complexity.
After that, the material system does, in principle, produce solutions to instances almost instantaneously.
In a sense, this evolutionary process can be regarded as a kind of preprocessing, but different from the preprocessing that is used in classical computation to decrease the size of the instances, an important concept in the domain of FPT.
Clearly, there are issues with scalability involved in EiM. It is likely that a limited amount of material, together with a limited amount of input and output channels, and a limited amount of configuration signals, has a bounded capability of
  solving instances of a  problem with an unbounded number of possible instances. It seems difficult to take all of these aspects into account in order to define a good measure for the capability of EiM to tackle hard problems.
Such problems are perhaps not the best candidates for the EIM approach. Instead, it might be better to focus future research on computational tasks that are difficult to accomplish with classical computational devices; not difficult in the sense of computational complexity but in the sense of developing and implementing the necessary computer programs to perform the tasks.
One might think of classification tasks like speech, face and pattern recognition.

\subsection{Optical computing}

It is possible to use optical systems to compute with photons rather than electrons.
Although this is a somewhat unconventional computational substrate, the computation performed is purely classical.

Some authors suggest more unconventional applications of optical components. \citet{Woods2005,Woods2009} discuss a form of spatial optical computing that encodes data as images, and computes by transforming the images through optical operations.
These include both analogue and digital encodings of data.

\citet{Reif1994} describe a particular idealised ray tracing problem cast as a decision problem, and show that it is Turing-uncomputable.  They also note that the idealisations do not hold in the physical world:
\begin{quote}
Theoretically, these optical systems can be viewed as general optical computing machines, if our constructions could be carried out with infinite precision, or perfect accuracy. However, these systems are not practical, since the above assumptions do not hold in the physical world. Specifically, since the wavelength
of light is finite, the wave property of light, namely diffraction, makes the theory of geometrical optics fail at the wavelength level of distances.
\end{quote}
\citet{Blakey2014} has furthermore formalised the intuition that the claimed hypercomputational power of even such idealised computers in fact vanishes once precision is properly considered.

\citet{Wu2014} construct an optical network than can act as an oracle for the Hamiltonian path decision problem.  Their encoding approach addresses the usual precision issue by having exponentially large delays; hence, as they say, it does not reduce the complexity of the problem, still requiring exponential time.
They do however claim that it can provide a substantial speed-up factor over traditional algorithms, and demonstrate this on a small example network. However, since the approach explores all possible paths in parallel, this implies an exponential power requirement, too.

\subsection{MEM-computing}

MEM-computing has been introduced by  \citet{Traversa2015a,Traversa2017,Traversa2015-UMM} as a novel non-Turing model of computation that uses interacting computational memory cells -- memprocessors -- to store and process information in parallel on the same physical platform, using the topology of the interaction network to form the specific computation. 

\citet{Traversa2015-UMM} introduce the Universal Mem\-computing Machine (UMM), and make a strong claim:
\begin{quote}
We also demonstrate that a UMM has the same computational power
as a non-deterministic Turing machine, namely it can solve \NP-complete problems in polynomial time. However, by virtue of its information overhead, a UMM needs only an amount of memory cells (memprocessors) that grows polynomially with the problem size. \ldots\ Even though these results do not prove the statement \NP\ = \Poly\ within the Turing paradigm, the practical realization of these UMMs would represent a paradigm shift from present von Neumann architectures bringing us closer to brain-like neural computation.
\end{quote}

The UMM is essentially an analogue device.
\citet{Traversa2015-UMM} state:
\begin{quote}
a UMM can operate, in principle, on an infinite number of continuous states, even if the number of
memprocessors is finite. The reason being that each memprocessor
is essentially an analog device with a continuous set of
state values
\end{quote}
then in a footnote acknowledge:
\begin{quote}
Of course, the actual implementation of a UMM will limit this continuous range to a discrete set of states whose density depends on the experimental resolution of the writing and reading operations.
\end{quote}

\citet{Traversa2015a} present an experimental demonstration with 6 memprocessors solving a small instance of the NP-complete version of the subset sum problem in only one step.  The number of memprocessors in the given design scales linearly with the size of the problem.  The authors state: 
\begin{quote}
the particular machine presented here is eventually limited by noise---and will thus require error-correcting codes to scale to an arbitrary number of memprocessors
\end{quote}

Again, the issue is an analogue encoding, here requiring a Fourier transform of an exponential number of frequencies, and accompanying exponential requirements on precision, and possibly power.
As we have emphasised earlier (\S\ref{sec:measure}), complexity is an asymptotic property, and small problem instances do not provide evidence of asymptotic behaviour.
See also \citet{Saunders2016} for a further critique of this issue.
\citet{Traversa2015a} state that this demonstration experiment ``represents the first proof of concept of a machine capable of working with the collective state of interacting memory cells'', exploring the exponentially large solution space in parallel using waves of different frequencies. 
\citet{Traversa2015a} suggest that this is similar to what quantum computing does when solving difficult problems such as factorisation. 
However, although quantum computing is a powerful and physical model, there is no evidence that it can efficiently solve \NP-hard problems. The power of quantum computing is quantum superposition and entanglement in Hilbert space, not merely classical wave computing using collective coherent classical states. The  oracles needed for a quantum computer to solve problems in \QMA\ most likely do not exist in the physical world. 
\citet{Aaronson2005} examines and refutes many claims to solving \NP-hard problems, demonstrating the different kinds of smuggling that are used, sweeping the eventual need for exponential resources under (very large) carpets. This is also underlined in  \citet{Aaronson2015a}, a blog post providing an extended critique of the claims in \citet{Traversa2015a}.

In order to address these scaling limitations of the analogue UMM, \citet{Traversa2017} present their \textit{Digital} Memcomputing Machine (DMM).  They claim that DMMs also have the computational power of nondeterministic Turing machines, able to solve \NP-complete problems in polynomial time with resources that scale polynomially with the input size.  They define a dynamical systems model of the DMM, and prove several properties of the model.  They provide the results of several numerical simulations of a DMM circuit solving small instances of an \NP-complete problem, and include a discussion of how their results suggest, though do not prove, that \Poly\ = \NP.  The computational power of the DMM is claimed to arise from its ``intrinsic parallelism'' of operation.  This intrinsic parallelism is a consequence of the DMM components communicating with each other in an analogue manner during a computational state transition step.  The DMM is digital insofar as it has a digital state at the beginning and end of a computational step.  However, its operation has the same fundamentally analogue nature as the UMM during the intrinsically parallel execution of a computational step, and so all the issues of precision and noise will still need to be addressed before any super-Turing properties can be established.

\subsection{Brain computing}

The brain is generally considered to be a natural physical adaptive information processor subject to physical law; 
see, e.g., \citet{Bassett2011,Jubat2015,Schaul2011,ChesiMoro2014}. 
As such it must essentially be a classical analogue ``machine'', and then the 
ECT states that it can, in principle, be simulated efficiently by a digital classical computer.

This ``limitation'' of the brain is not always accepted, however, especially by philosophers. Leaving aside far-fetched quantum-inspired models, brain-inspired models sometimes assume that the brain is able to efficiently solve \NP-hard problems, see e.g. \citet{Traversa2015a}, and therefore can serve as a model for super-Turing computing beyond classical digital DTMs.
The underlying idea is likely that (human) intelligence and consciousness are so dependent on processing power that classical digital computers cannot efficiently model an intelligent self-conscious brain -- such problems must necessarily be \NP-hard.

The view of this chapter's authors is that our brain (actually any animal brain) is a powerful but classical information processor, and that its workings remain to be found out. The fact that we have no real understanding of the difference between a conscious and unconscious brain is most likely not to be linked to the lack of processing capacity.

One side of the problem-solving capacity of the brain is demonstrated in game playing, 
as illustrated by the performance of the AlphaGo adaptive program of 
Google DeepMind \citep{Silver2016},  run on a digital computer and winning over both the European and the world champions \citep{Wired2016}, beating the human world champion 4--1 in a series of Go games. 
AlphaGo is adaptive, based on deep neural networks (machine learning) and tree search, probably representing a significant step toward powerful artificial intelligence  (AI). Since the human champion no doubt can be called intelligent, there is some foundation for ascribing some intelligence to the AlphaGo adaptive program.  The problem of characterising
artificial intelligence can be investigated via game playing \citep{Schaul2011}.

Games present \NP-hard problems when scaled up \citep{Viglietta2012}.
\citet{Aloupis2012} have proved the \NP-hardness of five of Nintendo's largest video game franchises. In addition, they prove \PSPACE-completeness of the Donkey Kong Country games and several Legend of Zelda games.  

For AlphaGo not only to show some intelligence, but also be aware of it, i.e., aware of itself, is of course quite a different thing. This does not necessarily mean that consciousness presents a dramatically harder computational task. But it then needs  mechanisms for self-observation and at least short-term memory for storing those observations. When AlphaGo then starts describing why it is making the various moves, thinking about them (!), and even recognising its errors, then one may perhaps grant it some consciousness, and perhaps even intuition. 

The argument that the brain violates the ECT appears to rest on the fact that brains can solve certain problems that are uncomputable or \NP-hard.  However, this argument confuses the \textit{class} of problems that have these properties with the individual instances that we can solve. So, for example, despite our being able to prove whether \textit{certain} programs terminate, there is no evidence that we can prove whether \textit{any} program terminates. 

We hold that the brain is a natural, physical, basically analogue information processor that \textit{cannot} solve difficult instances of  \NP-hard problems. Therefore intelligence, consciousness, intuition, etc. must be the result of natural computation processes that, in principle, can be {\em efficiently} simulated by a classical digital computer, given some models for the relevant brain circuitry. The question of polynomial overhead is a different issue. Perhaps it will be advantageous, or even necessary, to emulate some brain circuits in hardware in order to get reasonable response times of artificial brain models. Such energy-efficient neuromorphic hardware is already emerging \citep{Merolla2014,Esser2016} and may soon match human recognition capabilities \citep{Maass2016}.

\subsection{Conclusion}

These discussed forms of unconventional physically realisable computing are unable to solve \NP-hard problems in polynomial time, unless given access to an uncomputable oracle or some ``smuggled'' exponential resource, such as precision, material, or power. So, hard instances of \NP\ problems cannot be solved efficiently. Various approaches to solving \NP-hard problems result in at best polynomial speed-up. The quantum class \BQP\ does indicate that quantum computers can offer efficiency improvements for some problems outside the \NP-complete class (always assuming \Poly\ $\neq$ \NP).

\section{Overall Conclusions and Perspectives}

We have provided an overview of certain claims of hypercomputation and super-Turing computation in a range of unconventional computing devices.  Our over\-view covers many specific proposals, but is not comprehensive.
Nevertheless, a common theme emerges:
all the devices seem to rely on one or another \textit{unphysical} property to work: infinite times or speeds, or infinite precision, or uncomputable advice, or some unconsidered exponential physical resource.

There is value in examining a wide range of unconventional theoretical models of computation, even if those models turn out to be unphysical.
After all, even the theoretical model that is the DTM is unphysical: its unbounded memory tape, necessary for its theoretical power, cannot be implemented in our finite bounded universe.  Our physical digital computers are all finite state machines.
Claims about \textit{physical} implementability of models more powerful than the DTM, when the DTM itself is unphysical, need to be scrutinised carefully,
and not in the realm of mathematics (their theoretical power), but rather in that of physics (their implementable power).

One issue that UCOMP helps foreground is this existence of \textit{physical} constraints on \textit{computational} power  \citep{Aaronson2005,Potgieter2006,Denef2007,Aaronson2013,Aaronson2015a,Aaronson2009}. 
That there might be such physical limits to computational power may be difficult to accept, judging from a wealth of publications discussing and describing how to efficiently solve \NP-hard problems with physical computers. 
However, as we have argued in this chapter, there is no convincing evidence that classical (non-quantum) computational devices (whether analogue or digital, engineered or evolved) can be built to \textit{efficiently} solve problems outside classical complexity class \Poly.

The Laws of Thermodynamics, which address energy conservation and entropy increase, express bounds on what is physically possible.
A consequence of these laws is that perpetual motion machines are physically \textit{impossible}, and any purported design will have a flaw somewhere.
These are laws of physics, not provable mathematical theorems, but are well-evidenced.
The laws are flexible enough that newly discovered phenomena (such as the convertibility of mass and energy) can be accommodated.

The CTT and ECT appear to play an analogous role in computation.
They express bounds on what computation is \textit{physically} possible.
A consequence of these bounds, if they are true, is that hypercomputing and super-Turing computing are \textit{physically} impossible.
And they are similarly flexible that newly discovered phenomena (such as quantum computing) can be accommodated.
This demonstrates a fundamental and deep connection between computation and the laws of physics: computation can be considered as a natural science, constrained by reality, not as an abstract branch of mathematics \citep{Horsman2017}.

The CTT and ECT can be expressed in a, slightly tongue-in-cheek, form echoing the laws of thermodynamics:

\begin{description}
\item{\it 1st Law of Computing:} You cannot solve uncomputable or \NP-hard problems efficiently unless you have a physical infinity or an efficient oracle.
\item{\it 2nd Law of Computing:} There are no physical infinities or efficient oracles.
\item{\it 3rd Law of Computing:} Nature is physical and does not solve uncomputable or \NP-hard problems efficiently.
\item{\it Corollary:} Nature necessarily solves uncomputable or \NP-hard problems only approximately.
\end{description}

This raises the question: what can UCOMP do? Given that UCOMP does not solve uncomputable or even \NP-hard problems (and this also applies to quantum computing), what is the future for UCOMP?
Instead of simply trying to do conventional things faster, UCOMP can focus on novel applications, and novel insights into computation, including:
\begin{itemize}
\item Insight into the relationship between physics and computation
\item New means for analogue simulation/optimisation
\item Huge parallelisation
\item Significant polynomial speed-up
\item Novel forms of approximate solutions
\item Cost-effective solutions
\item Solutions beyond the practical capability of digital HPC
\item Solutions in novel physical devices, for example, programmed synthetic biological cells
\end{itemize}

UCOMP offers many things.  But it does not offer hypercomputing or super-Turing computing realisable in the physical world.

\subsection*{Acknowledgements}
H.B. acknowledges funding from the European Community's Seventh Framework Programme (FP7/2007-2013) under grant agreement number 317662 (the FP7 FET NASCENCE project).
S.S. acknowledges partial funding by the EU FP7 FET Coordination Activity TRUCE (Training and Research in Unconventional Computation in Europe), project reference number 318235.
G.W.\ acknowledges support by the European Commission through the FP7 SYMONE and Horizon 2020 RECORD-IT projects,
and by Chalmers University of Technology.

\bibliographystyle{dcu}
\bibliography{goran,susan}
\end{document}